\documentstyle[12pt,graphicx]{article}

\begin{document}

\title{Synchrotron Self Absorption in GRB Afterglow}
\author{Jonathan Granot, Tsvi Piran and  Re'em Sari \\ \vspace{-0.5cm} \\
\small{Racah Institute, Hebrew University, Jerusalem 91904, Israel}}
\maketitle
\date{}

\begin{abstract}
  GRB afterglow is reasonably described by synchrotron emission from
  relativistic blast waves at cosmological distances. We perform
  detailed calculations taking into account the effect of synchrotron
  self absorption. We consider emission from the whole region behind
  the shock front, and use the Blandford McKee self similar solution
  to describe the fluid behind the shock. We calculate the spectra and
  the observed image of a GRB afterglow near the self absorption
  frequency $\nu_a$ and derive an accurate expression for $\nu_a$.  We
  show that the image is rather homogeneous for $\nu<\nu_a$, as
  opposed to the bright ring at the outer edge and dim center, which
  appear at higher frequencies. We compare the spectra we obtain to
  radio observations of GRB970508. We combine the calculations of the
  spectra near the self absorption frequency with other parts of the
  spectra and obtain revised estimates for the physical parameters of
  the burst: $E_{52}=0.53$, $\epsilon_e=0.57$, $\epsilon_B=0.0082$,
  $n_1=5.3$. These estimates are different by up to two orders of
  magnitude than the estimates based on an approximate spectrum.
\end{abstract}


\section{Introduction}
The detection of delayed x-ray, optical and radio emission following a
GRB, known as GRB afterglow, is described reasonably well by emission
from a spherical relativistic shell, decelerating upon collision with
an ambient medium (Waxman 1997a, Wijers, Rees \& M\'esz\'aros 1997,
Katz \& Piran 1997, Sari, Piran \& Narayan 1998). A relativistic blast
wave expands through the ambient medium, continuously heating up fresh
matter as it passes through the shock. In these models, the GRB
afterglow is the result of synchrotron emission of the relativistic
electrons of the heated matter.

Several recent works considered emission from various regions on or
behind the shock front (Waxman 1997c, Sari 1998, Panaitescu \&
M\'esz\'aros 1998, Granot, Piran \& Sari 1998 (GPS hereafter),
Gruzinov \& Waxman 1998). These authors considered the spectra near
the peak frequency $\nu_m$ and found that exact calculations of the
spectrum could differ by up to one order of magnitude from simpler
estimates.

We consider emission from an adiabatic highly relativistic blast wave
expanding into a cold and uniform medium. We consider the effect of
the whole volume behind the shock front, the importance of which was
stressed in GPS. The hydrodynamics is described by the Blandford McKee
(1976 denoted BM hereafter) self similar solution.  We consider
synchrotron emission and we ignore Compton scattering and electron
cooling. Similar to GPS, we consider several models for the evolution
of the magnetic field.

Synchrotron self absorption becomes significant below a critical
frequency $\nu_a$ called the self absorption frequency.  We assume
that $\nu_a \ll \nu_m$, where $\nu_m$ is the peak frequency, which is
reasonable for the first few months after the burst. While the
spectrum near $\nu_m$ was quite extensively studied, so far only order
of magnitude estimates of the spectrum near the self absorption
frequency $\nu_a$ were done.  

For a system in which $\nu_a \ll \nu_m$, the spectrum, $F_{\nu}$, is
proportional to $\nu^2$ for $\nu \ll \nu_a$, rather than the standard
$\nu^{5/2}$, because almost all the low frequency radiation is emitted
by electrons with a typical synchrotron frequency much higher than
$\nu_a$ (Katz 1994).  The spectra far above $\nu_a$ is
proportional to $\nu^{1/3}$. In this paper we explore the spectra near
$\nu_a$, and find how the two asymptotic forms join together.  An
analysis of the spectra over a wider range of frequencies was made by
Sari, Piran \& Narayan (1998), and a detailed analysis of the spectra
near the peak frequency, taking a full account of the BM solution, was
made in GPS and Gruzinov \& Waxman (1998).

The physical model is described in \S 2. In \S 3 we describe the
computational formalism. The spectra for several magnetic field models
and the observed image of a GRB afterglow at various frequencies are
presented in \S 4.  In \S 5 we compare the calculated spectra to radio
observations of the afterglow of GRB970508. When we use the modified
calculations of the self absorption spectra and of the spectra around
the peak (GPS) and the cooling frequency (Sari, Piran \& Narayan 1998)
we find new estimates of the parameters of GRB970508. These estimates
are different by more than an order of magnitude than estimates based
on a simpler broken power law spectra (Wijers \& Galama 1998).

\section{The Physical Model}
The underlying model assumes an ultra-relativistic spherical blast
wave expanding into a cold and uniform medium. The blast wave
constantly heats fresh matter, and the observed afterglow is the
result of synchrotron emission of the relativistic electrons of the
heated matter. We consider an adiabatic evolution, where the fluid
behind the shock is described by the BM self similar solution. It has
been numerically verified that for an adiabatic evolution with general
initial conditions the solution approaches the BM solution (Kobayashi,
Piran \& Sari 1998). The BM solution is expressed in terms of the
similarity variable $\chi$, which is defined by:
\begin{equation}
\label{def_chi}
\chi \equiv 1+16 \gamma_f^2\left({R-r \over R}\right) \ ,
\end{equation}
where $R$ is the radius of the shock front, $r$ is the distance of a
point from the center of the burst and $\gamma_f$ is the Lorentz
factor of the matter just behind the shock. The BM solution is given
by:
\begin{equation}
\label{BM}
n'= 4 \gamma_f n_0 \chi^{-5/4} \ \ \ ,\ \ \ 
\gamma=\gamma_f \chi^{-1/2} \ \ \ ,\ \ \ 
e'=4 n_0 m_p c^2 \gamma_f^2 \chi^{-17/12} \ ,
\end{equation}
where $n'$ and $e'$ are the number density and the energy density in
the local frame, respectively, $\gamma$ is the Lorentz factor of the
bulk motion of the matter behind the shock, $m_p$ is the mass of a
proton and $n_0=n_1\times 1\rm cm^{-3}$ is the proper number density
of the unshocked ambient medium.

We consider three alternative models for the magnetic field: $B$,
$B_{\perp}$ and $B_{rad}$. $B$ satisfies $e'_{B'}=\epsilon_B e'$ (i.e.
equipartition) everywhere. For the two other magnetic field models we
assume they acquire $e'_{B'}=\epsilon_B e'$ on the shock front, and
from then on, evolve according to the ``frozen field'' approximation.
This implies that the magnetic fields equal $B \chi^{3\delta/2}$, where
$\delta = 7/18$ when the magnetic field is in the radial direction
($B_{rad}$) and $\delta = -7/36$ when the magnetic field is
perpendicular to the radial direction ($B_{\perp}$) (see GPS).

We assume that the energy of the electrons is everywhere a constant
fraction of the internal energy: $e'_{el}=\epsilon_e e'$, and that the
shock produces a power law electron distribution:
$N(\gamma_e)=K{\gamma_e}^{-p}$ \footnote{for the energy of the
electrons to remain finite we must have $p>2$.}  for $\gamma_e \ge
\gamma_{min}$. We use $p=2.5$ wherever  a definite numerical value of
$p$ is needed. The constants $K$ and $\gamma_{min}$ in the electron
distribution can be calculated from the number density and energy
density:
\begin{equation}
\label{el dis}
\gamma_{min}=\left({p-2\over p-1}\right){\epsilon_e e' \over n'm_ec^2} 
\ \ \ \ \ \ ,\ \ \ \ \ \ 
K=(p-1) n' \gamma_{min}^{p-1} \ .
\end{equation}

We consider frequencies which are much lower than the typical
synchrotron frequency $\nu \ll \nu_{syn}$, where the spectral power of
an electron emitting synchrotron radiation can be approximated by:
\begin{equation}
\label{P_enu}
P'_{\nu',e}\cong{2^{5/3}\pi\over\Gamma(1/3)}{q_e^3 B'\sin\alpha \over m_e c^2}
\left({\nu' \over \nu'_{syn}}\right)^{1/3} \quad , \quad
\nu'_{syn} \equiv {3 \gamma_e^2 q_e B' \sin \alpha \over 4 \pi m_e c} \ ,
\end{equation}
(Rybicki \& Lightman 1979) where $\Gamma$ is the gamma function, $B'$
is the magnetic field, $m_e$ and $q_e$ are the mass and the electric
charge of the electron, respectively, and $\alpha$ is the angle
between the directions of the electron's velocity and the magnetic
field, in the local frame.

$R_l$ and $\gamma_l$ are the location of the shock and its Lorentz
factor on this line of sight and $R_l/\gamma_l$ is an estimate of the
shell thickness in the local frame. The optical depth along the LOS
can be approximated by $\alpha'_{\nu'}R_l/\gamma_l$ where the
absorption coefficient $\alpha'_{\nu'}$ is taken at $y=\chi=1$. The
``back of the envelope'' estimate for the self absorption frequency
$\nu_0$ is therefore the frequency which satisfies:
$\alpha'_{\nu'_0}R_l/\gamma_l=1$. $\nu_0$ is given by:
\begin{equation}
\label{nu_0} 
\nu_0=4.24 \times 10^{9} (1+z)^{-1} \left({p+2 \over 3p+2}\right)^{3/5}
{(p-1)^{8/5} \over (p-2)} \epsilon_e^{-1} \epsilon_B^{1/5} E_{52}^{1/5}
n_1^{3/5} \rm Hz \ ,
\end{equation}
where $z$ is the cosmological red shift and $E=E_{52}\times10^{52}\rm
ergs$ is the total energy of the shell. We also define a ``standard''
flux density $F_0$, which is an approximate expression for the flux
density at the self absorption frequency.  The peak flux and peak
frequency can be approximated by:
\begin{equation}
\label{F_m&nu_m}
F_m\equiv{(1+z)\over 3}{n_1R_l^3\over d_L^2}\left({\gamma P'_e\over 
\nu'_{syn}}\right)(y=\chi=1) \quad, \quad 
\nu_m\equiv\nu_{syn}(\gamma_{min},y=\chi=1) \ ,
\end{equation}
(see $F_0$ and $\nu_T$ in GPS) where
$P'_e=(4/3)\sigma_Tc\gamma_e^2e'_B$ is the total emitted power of an
extreme relativistic electron (Rybicki \& Lightman 1979), $\sigma_T$
is the Thomsom cross section and $d_L=d_{L28}\times 10^{28}\rm cm$ is
the luminosity distance \footnote{Here and latter we use a general
  cosmological model.  For a given model we should substitute the
  appropriate expression for $d_L(1+z,H_0)$.}.  Now we can define:
$F_0\equiv F_m(\nu_a/\nu_m)^{1/3}$:
\begin{equation}
\label{F_0}
F_0=1.31(1+z) \left({p+2 \over 3p+2}\right)^{1/5}{(p-1)^{6/5} \over (p-2)}
d_{L28}^{-2}\epsilon_e^{-1} \epsilon_B^{2/5} E_{52}^{9/10} n_1^{7/10}
T_{days}^{1/2} \rm mJy \ ,
\end{equation}
where $T_{days}$ is the observed time in days.

We express the observed frequency $\nu$ in units of $\nu_0$: $\nu
\equiv \phi \nu_0$, thus introducing the dimensionless variable $\phi$
which we use to express our results.

\section{The Formalism}
We consider a system that is moving relativistically while emitting
radiation, and obtain a formula for the flux density measured by a
distant observer, allowing for self absorption. We denote quantities
measured in the local rest frame of the matter with a prime, while
quantities without a prime are measured in the observer frame.

The specific intensity $I_{\nu}$ (energy per unit time per unit area
per unit frequency per unit solid angle) satisfies the radiative
transfer equation:
\begin{equation}
\label{rte}
{dI_\nu \over ds} = j_{\nu} - \alpha_{\nu}I_{\nu} \ ,
\end{equation}
where $j_{\nu}$ and $\alpha_{\nu}$ are the emission coefficient and
the absorption coefficient, respectively, and $s$ is the distance
along the beam.  When self absorption becomes important the
contribution to $I_{\nu}$ depends on the optical depth along the path
within the emitting system and $I_{\nu}$ should be integrated
separately for every trajectory.

For the BM solution, the radius of the shock front and the Lorentz
factor of the matter just behind the shock, on the LOS for a given
observed time $T$, are given by:
\begin{equation}
\label{Rl_Gl} 
\gamma_l \cong 3.65E_{52}^{1/8}n_1^{-1/8}T_{days}^{-3/8} \quad, \quad
R_l \cong 5.53\times 10^{17}E_{52}^{1/4}n_1^{-1/4}T_{days}^{1/4}\rm{cm} \ .
\end{equation}
For a given observed time $T$, we use the coordinates $y$ and $\chi$,
instead of the polar coordinates $r$ and $\theta$, where $y$ is
defined by $y \equiv R/R_l$. $R$ is the radius of the shock front at
the time $t$ at which a photon must be emitted in order to reach the
observer at the observed time $T$. A photon emitted at a location
$(r,\theta)$ at a time $t$ in the observer frame, reaches the observer
at an observed time $T$ given by:
\begin{equation}
\label{time} 
T=t-{r \mu \over c} \ ,
\end{equation}
where $\mu \equiv \cos \theta$ (for more details on the coordinates,
see GPS).

\begin{figure}
\centering
\noindent
\includegraphics[width=9.8cm]{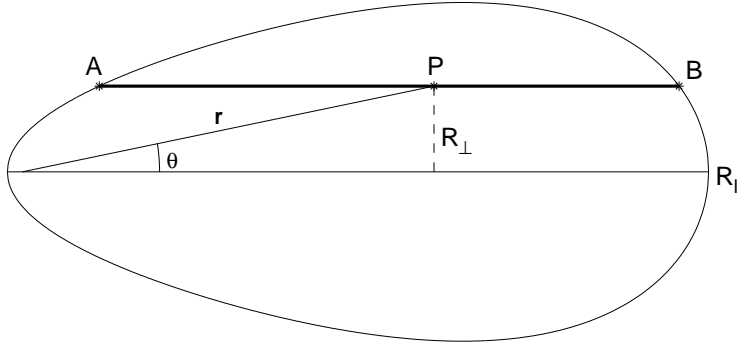}
\caption{\label{Fig1}The egg-Shaped curve is the boundary of the
  region from which photons reach a distant observer simultaneously.
  The observer is located far to the right and the symmetry axis is
  the LOS from the source of the burst to the observer. The bold line
  represents the trajectory of a photon that reaches the observer.
  For a distant observer these trajectories are almost parallel to the
  LOS, and are therefore characterized by their distance $R_{\perp}$ from
  the LOS. $I_{\nu}=0$ at point A, and reaches its final value at
  point B. A photon emitted at a point P can be absorbed or cause
  stimulated emission at any point along the trajectory, until it
  passes the shock front at point B. }
\end{figure}

Since the observer is distant, the trajectory of a photon that reaches
the observer is almost parallel to the LOS in the observer frame (see
Figure \ref{Fig1}). We can therefore parameterize the various
trajectories by their distance from the LOS. The distance $R_{\perp}$
of a point from the LOS is given by:
\begin{equation}
\label{Rp} 
R_{\perp} \equiv r\sin \theta \cong R_l y \sqrt {1-\mu^2}
\cong {\sqrt2 R_l \over 4\gamma_l} \sqrt {y-\chi y^5} \ .
\end{equation}
The maximal value of $R_{\perp}$ is obtained for $\chi=1$,
$y=5^{-1/4}$ and is given by:
\begin{equation}
\label{Rpmax} 
R_{\perp,max}= 3.91\times10^{16}E_{52}^{1/8}n_1^{-1/8}T_{days}^{5/8}\rm{cm}\ ,
\end{equation}
We express $R_{\perp}$ in units of $R_{\perp,max}$, introducing the
dimensionless variable:
\begin{equation}
\label{x} 
x \equiv {R_{\perp} \over R_{\perp,max}} = {5^{5/8} \over 2} 
\sqrt {y-\chi y^5} \ .
\end{equation}
In order to solve equation \ref{rte}, explicit expressions for the
absorption and the emission coefficients are needed. We consider an
isotropic electron velocity distribution, and since $P'_{\nu',e}\propto
\sin^{2/3} \alpha$ we use the averaged value $\langle\sin^{2/3}
\alpha\rangle = \sqrt\pi \Gamma(1/3)/5\Gamma(5/6)$.  The total power
per unit volume per unit frequency in the local frame is given by:
\begin{equation}
\label{P_nu}
P'_{\nu'}=\int_{\gamma_{min}}^{\infty}d\gamma_e N(\gamma_e) P'_{\nu',e} =
{64\pi^{13/6}3^{2/3}\over5\Gamma(5/6)}{(p-1)^{5/3}\over(3p-1)(p-2)^{2/3}}
{\epsilon_B^{1/3}\gamma_ln_0^{4/3}q_e^{8/3}\nu'^{1/3}\over m_p^{1/3}c
y^{3/2}\chi^{29/18-\delta}} \ .
\end{equation}

Assuming the emission and absorption are isotropic in the local rest
frame of the matter, $j'_{\nu'}=P'_{\nu'}/4\pi$ and the absorption
coefficient is given by:
\begin{equation}
\label{alpha_nu} 
\alpha'_{\nu'} = {(p+2) \over 8 \pi m_e \nu'^2}\int^{\infty}_{\gamma_{min}} 
d\gamma_e P'_{\nu',e}(\gamma_e){N(\gamma_e) \over \gamma_e}
=\end{equation}
$$
{8 \pi^{7/6}3^{2/3}\over5\Gamma(5/6)}{(p+2)(p-1)^{8/3}\over(3p-2)(p-2)^{5/3}} 
{\epsilon_b^{1/3}n_0^{4/3}q_e^{8/3}\over\epsilon_e^{5/3}m_p^{4/3}c\nu'^{5/3}
\chi^{13/9-\delta}} \ ,
$$
(Rybicki \& Lightman 1979). Keeping in mind that $j_{\nu}/\nu^2$ and
$\alpha_{\nu}\nu$ are Lorentz invariant and
$\nu'=\nu\gamma(1-\beta\mu)$, we can write the radiative transfer
equation (equation \ref{rte}) explicitly.  We first write an
expression for the optical depth to the observer, which will later
help us gain some intuition for the results. The optical depth to the
observer is given by:
\begin{equation}
\label{tao} 
\tau(x,y)=\int\alpha_{\nu}ds={2\over\phi^{5/3}}\int_{y}^{y_{max}(x)}
dy{y^{5/3}\over\chi^{10/9-\delta}(x,y)(1+7\chi(x,y)y^4)^{2/3}} \ ,
\end{equation}
where $\chi(x,y)$ is obtained from equation \ref{x}, and $y_{max}(x)$
is obtained by solving $\chi(x,y)=1$ for $y$.

We define a ``typical'' specific intensity $I_0$ by: $I_0 \equiv
S_{\nu}(y=\chi=1)$, where the source function $S_{\nu}$ is defined as:
$S_{\nu} \equiv j_{\nu}/\alpha_{\nu}$. Expressing $I_{\nu}$ in units
of $I_0$: $I_{\nu} \equiv \tilde I_{\nu} I_0$, we write equation
\ref{rte} in terms of the dimensionless variables $\tilde I_{\nu}$:
\begin{equation}
\label{rte2}
{d\tilde I_\nu \over dy} = 
{2 y^{5/3} \over \phi^{5/3} \chi^{10/9-\delta} (1-7\chi y^4)^{2/3}}
\left({8 y \chi^{1/3} \over (1-7\chi y^4)}-\tilde I_{\nu} \right) \ ,
\end{equation}
where we took $ds \cong R_l dy$. This equation can be solved
numerically for a given magnetic field model and a given value of
$\phi$, for different values of $x$, thus obtaining $I_{\nu}(x)$.  The
observed flux density is given by:
\begin{equation}
\label{flux}
F_{\nu}(T)\cong{(1+z)\over d_L^2}\int dS_{\perp}I_{\nu}=2\pi(1+z)
\left({R_{\perp,max}(T) \over d_L}\right)^2\int_0^1xdx I_{\nu}(x,T) \ .
\end{equation}
As can be seen from equation \ref{flux}, the surface brightness is
proportional to $I_{\nu}$: $dF_{\nu}/dS_{\perp} \propto I_{\nu}$. This
means that we also obtain the observed image of a GRB afterglow, by
calculating the observed flux density in this way.

\section{The Spectra and Observed Image}
\begin{figure}
\centering
\noindent
\includegraphics[width=13cm]{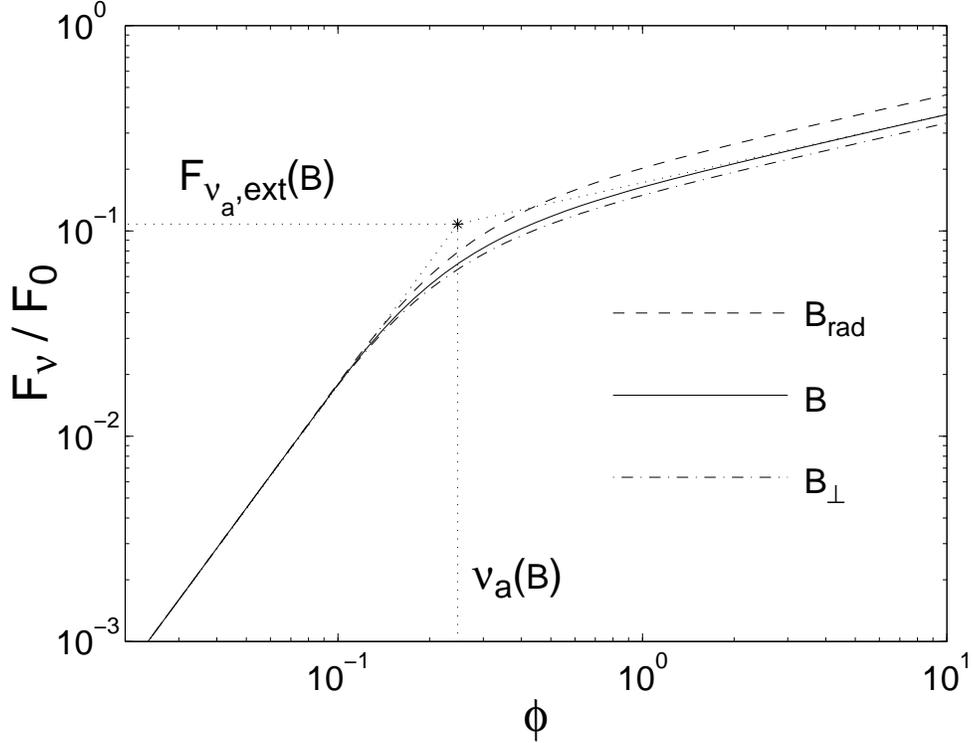}
\caption{\label{Fig2}The spectra for different magnetic field
  models. The frequency $\nu_a$ and the flux density $F_{\nu_a,ext}$
  are defined at the point where the extrapolations of the power laws
  meet, as is illustrated for the equipartition magnetic field model
  $B$. $\nu_a$ is constant in time, $F_{\nu_a,ext}\propto T^{1/2}$ and
  both don't change significantly between the different magnetic field
  models.  For $\phi \ll 1$ ($\nu \ll \nu_{a}$) the system is
  optically thick, and the flux density reflects the electron
  distribution (or the Lorentz boosted ``effective temperature'' of
  the electrons) and is independent of the magnetic field model.}
\end{figure}

Solving equations \ref{rte2} and \ref{flux}, we calculate the spectra
for the three magnetic field models ($B$, $B_{rad}$ and $B_\perp$).
These spectra are shown in Figure \ref{Fig2}. We define the self
absorption frequency $\nu_a$ as the frequency at which the asymptotic
high and low frequency power laws meet. For the equipartition magnetic
field model $B$ we obtain:
\begin{equation}
\label{nu_a}
\nu_a=0.247\nu_0=1.05 \times 10^{9} (1+z)^{-1} 
\left({p+2 \over 3p+2}\right)^{3/5}
{(p-1)^{8/5} \over p-2} \epsilon_e^{-1} \epsilon_B^{1/5} E_{52}^{1/5}
n_1^{3/5} \rm Hz \ ,
\end{equation}
while for the $B_{\perp}$ model it is lower by $6\%$ and for the
$B_{rad}$ model it is higher by $14\%$. For $\nu \ll \nu_a$ the system
is optically thick , and therefore the radiation reaching an observer
is essentially emitted near the edge facing the observer. It reflects
the electron distribution there, and it is independent of the magnetic
field model. For $\nu > \nu_a$ the system is optically thin and the
flux density is different for the various magnetic field models.

\begin{figure}
\centering
\noindent
\includegraphics[width=9.4cm]{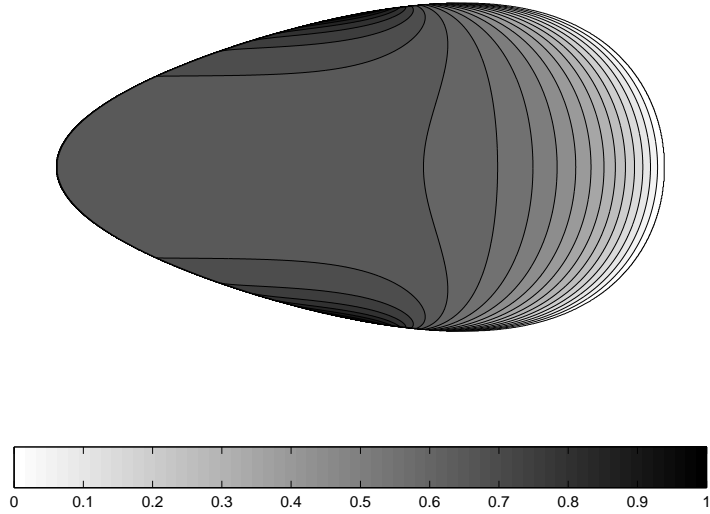}
\caption{\label{Fig3}The optical depth $\tau$ to the observer, divided
  by the maximal optical depth
  ($\tau=\tau_{max}=1.08\times(\nu/\nu_a)^{-5/3}$ is black).
  $\tau_{max}$ is obtained at $x\equiv R_{\perp}/R_{\perp,max}=0.93$.
  The contour lines are equally spaced with a $5\%$ interval between
  following contour lines.
    }
\end{figure}

The ratio $\tau/\tau_{max}$ (where $\tau$ is the optical depth and
$\tau_{max}$ is its maximal value) for the equipartition $B$ model is
shown in Figure \ref{Fig3}.  Since $\tau_{\nu} \propto \phi^{-5/3}$
everywhere, $\tau/\tau_{max}$ is frequency independent. The contour
lines of $\tau_{\nu}$ are dense where the absorption coefficient
$\alpha_{\nu}$ is large. $\tau_{max}$ is obtained at $x\cong 0.93$,
i.e. quite close to the edge of the image, since then the whole
trajectory to the observer is relatively close to the shock front,
implying large values of $\alpha_{\nu}$ and a large contribution to
$\tau_{\nu}$.  $\tau_{max}$ is a good indicator for the opacity of the
system, and for the equipartition $B$ model it is given by:
$\tau_{max}=1.08\times(\nu/\nu_a)^{-5/3}$.

We define $F_{\nu_a,ext}$ as the extrapolated flux density at $\nu_a$
(see Figure \ref{Fig2}).  For the equipartition $B$ model we obtain:
\begin{equation}
\label{F_extr}
F_{\nu_a,ext}=0.108F_0=142(1+z) 
\left({p+2 \over 3p+2}\right)^{1/5}{(p-1)^{6/5} \over (p-2)}
d_{L28}^{-2} \epsilon_e^{-1} 
\epsilon_B^{2/5} E_{52}^{9/10} n_1^{7/10}T_{days}^{1/2} \rm \mu Jy \ ,
\end{equation}
While for the $B_{\perp}$ model it is lower by $12\%$ and for the
$B_{rad}$ model it is higher by $23\%$. The actual flux density at
$\nu_a$ is around $35\%$ lower than $F_{\nu_a,ext}$: $F_{\nu_a} \cong
0.65 F_{\nu_a,ext}$. The values of $F_{\nu_a,ext}$ and $\nu_a$ are
useful, since for $\nu \ll \nu_a$ magnitude smaller than $\nu_a$
(assuming $\nu_m > \nu_a$) the flux density is given by: $F_{\nu}
\cong F_{\nu_a,ext}(\nu/\nu_a)^2$, and for $\nu_a \ll \nu \ll \nu_m$
it is given by: $F_{\nu} \cong F_{\nu_a,ext}(\nu/\nu_a)^{1/3}$. Both
approximations are already good within a few percent for frequencies 
a factor $\sim 3$ below or above $\nu_a$, respectively.

It is interesting to compare the spectra obtained for the BM solution
to that obtained for a simplistic model of a static homogeneous disk.
This comparison should help us learn whether the fact that the spectra
is rounded up near $\nu_a$ should be attributed mainly to the specific
hydrodynamics used, or whether it is a more general feature of a
calculation accounting for self absorption.  For a static disk we
obtain the well known result:
\begin{equation}
\label{disk}
F_{\nu}=F_{\nu_a,ext}^* \psi^2\left(1-\exp[{-\psi^{-5/3}}]\right) 
\quad , \quad \psi \equiv {\nu \over \nu_a^*}\ , 
\end{equation}
where the constants $F_{\nu_a,ext}^*$ and $\nu_a^*$ are determined by
the radius and width of the disk and by the hydrodynamic parameters of
the emitting matter within the disk. If these parameters are set so
that the two stared quantities equal those in equations \ref{F_extr}
and \ref{nu_a}, respectively, the resulting flux density is very
similar to that obtained for the BM solution. Substituting equations
\ref{F_extr} and \ref{nu_a} into equation \ref{disk} can thus serve as
a good approximation (better than $3\%$) for the observed flux density
at $\nu \ll \nu_m$. This similarity implies that the shape of the
spectrum near $\nu_a$ is rather independent of the specific
hydrodynamic solution considered. On the other hand, the exact values
of $\nu_a$ and $F_{\nu_a,ext}$ depend significantly on the
hydrodynamics, and can not be determined without the detailed
calculation. A simplified calculation for $\nu_a^*$ could for example
yield $\nu_0$ (equation \ref{nu_0}) instead of $\nu_a$ that is given in
equation \ref{nu_a}.

\begin{figure}
\centering
\noindent
\includegraphics[width=13cm]{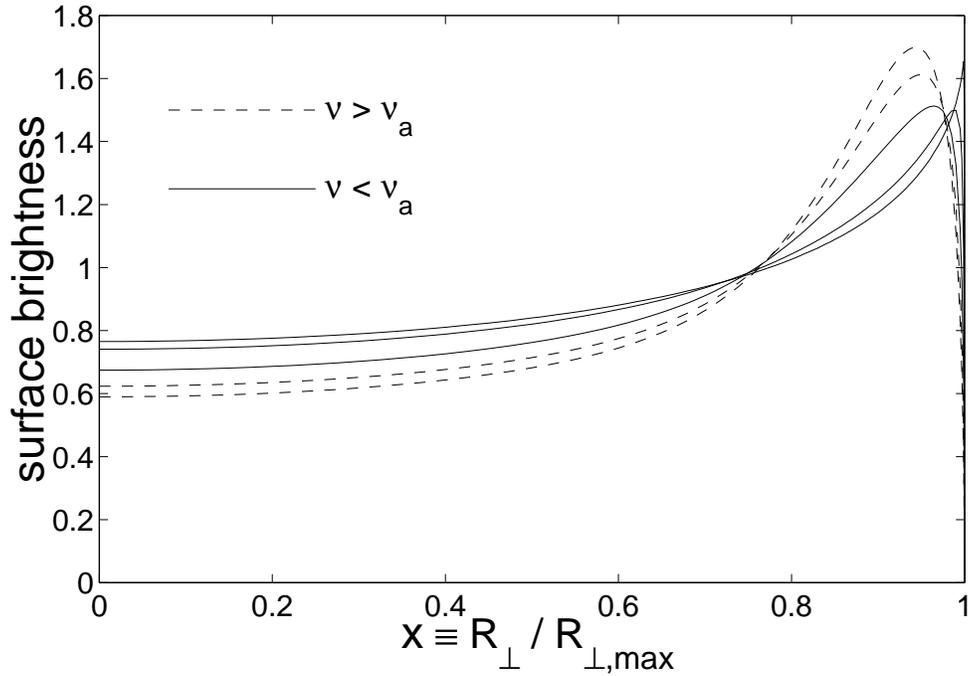}
\caption{\label{Fig4}The surface brightness divided by the
  average surface brightness, as a function of $x$, for
  $Log_{10}(\phi)=-1.5,-1 -0.75,-0.5,0$ (corresponding to
  $\nu/\nu_a=0.13,0.40,0.72,0.28,4.05$). At high frequencies the
  contrast between the center and the edge of the image is larger than
  at low frequencies. At low frequencies the system is optically
  thick, and the surface brightness reflects the ``effective
  temperature'' of the electrons at the edge of ``egg'' depicted in
  Figure \ref{Fig1}, on the side facing the observer.}
\end{figure}

\begin{figure}
  \centering
\noindent
\includegraphics[width=13cm]{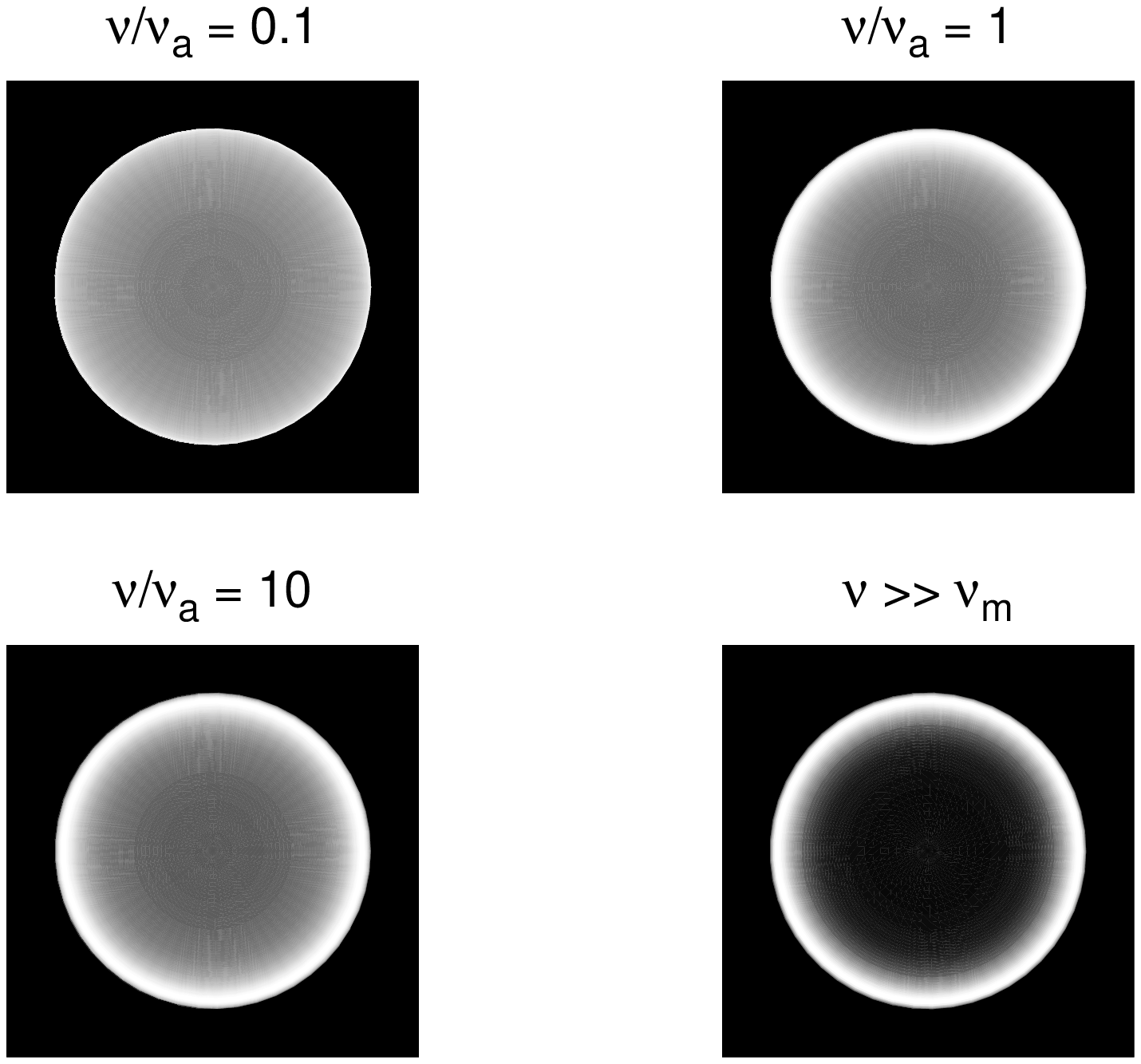}
\caption{\label{Fig5}The observed image of a GRB afterglow, at
  several frequencies. At high frequencies there is a bright ring near
  the outer edge of the image, and the contrast between the center and
  the edge of the image is larger than at low frequencies.  At low
  frequencies the surface brightness increases as one moves from the
  center towards the edge, until it drops very sharply, due to the
  fact that the system becomes optically thin near the edge. The last
  image, for $\nu \gg \nu_M$, is taken from GPS and it is brought as a
  reference, to illustrate the change in the relative surface
  brightness along the image over a large range of frequencies. At
  $\nu \gg \nu_m$ there is a thin bright ring at the outer edge of the
  image, and the surface brightness at the center is only a few
  percent of its maximal value.}
\end{figure}

The surface brightness as a function of $x \equiv
R_{\perp}/R_{\perp,max}$ for the equipartition magnetic field model
$B$ is shown in Figure \ref{Fig4}, for a few representative values of
$\phi\equiv\nu/\nu_0$. An illustration of the observed image of a GRB
afterglow, which is implied from the surface brightness, is shown in
Figure \ref{Fig5}. As a reference, the image for $\nu \gg \nu_m$,
which was derived in GPS, is also presented. 

For $\nu \gg \nu_a$ the system is optically thin and the result
coincides with that obtained in GPS for $\nu \ll \nu_m$.  There is a
bright ring near the outer edge of the image, and the surface
brightness at the center of the image is $58\%$ of its average value.
As $\nu$ decreases, the contrast between the edge and the center of
the image decrees and the system becomes increasingly optically thick.
As a result, the surface brightness reflects the Lorentz boosted
``effective temperature'' of the electron distribution, at the outer
edge of the ``egg'' depicted in Figures \ref{Fig1} and \ref{Fig3}, on
the side facing the observer.  Larger values of $x$ correspond to
smaller shock radii $R$, and since the electrons posses a larger
Lorentz factor (i.e. a larger ``effective temperature'') at smaller
radii, the surface brightness increases with $x$. This is true as long
as long as the optical depth is still large.  Since the length of the
trajectory within the system approaches zero as $x \to 1$, for every
given frequency $\nu$ the system becomes optically thin for values of
$x$ sufficiently close to $1$. For frequencies smaller than $\nu_a$ by
more than one or two orders of magnitude, this occurs at $1-x \ll 1$
so that the drop in the surface brightness near $x=1$ is extremely
sharp. For $\nu \ll \nu_a$ the surface brightness at the center of the
image is $77\%$ of its average value, resulting in an almost uniform
disk, rather than a ring which is obtained for $\nu > \nu_a$. The
uniformity of the image for $\nu \ll \nu_a$ stands out even more when
compared to $\nu \gg \nu_m$, where there is a thin bright ring on the
outer edge of the image and the surface brightness at the center is
only a few percent of its maximal value.

\section{Comparison to Observations}
We now fit the theoretical spectra calculated, to the radio
frequencies observations of the afterglow of GRB970508 (Fig. 4 of
Shepherd 1998). Perhaps a future near enough GRB will result in a
sufficient resolution, that will enable us compare the predicted
images of a GRB afterglow to observations.

\begin{figure}
\centering
\noindent
\includegraphics[width=9.9cm]{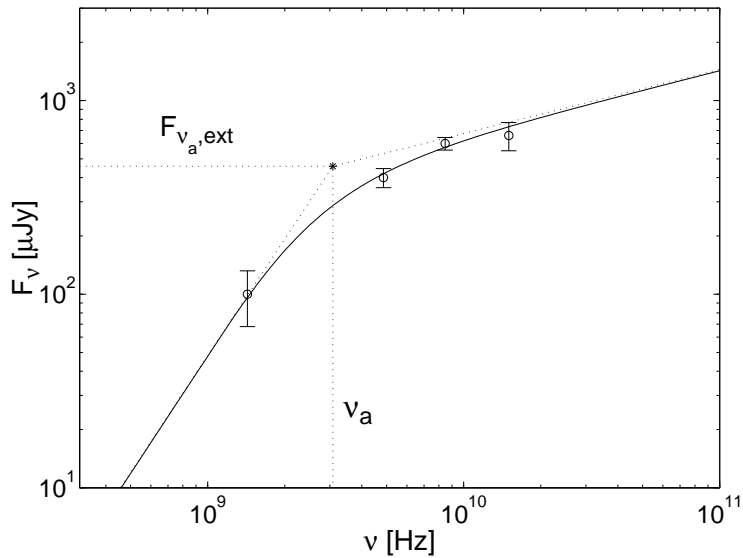}
\caption{\label{Fig6}A fit of the calculated spectra, to radio
  observations of the afterglow of GRB970508. We obtained: $\nu_a=3.1
  \pm 0.4 \times 10^9 \rm Hz$ and $F_{\nu_a,ext}= 450 \pm 37 \rm \mu
  Jy$.}
\end{figure}

Since both $\nu_0$ and $F_0$ depend on the parameters of the model
(see equations \ref{nu_0} and \ref{F_0}), we have two degrees of
freedom in trying to fit the calculated spectra to the observed data.
We fit the equipartition magnetic field model $B$ to the data in
Shepherd et al (1998), and obtain:
\begin{equation}
\label{constraints1}
\nu_a=3.1 \pm 0.4 \times 10^9 \rm Hz \quad,\quad 
F_{\nu_a,ext}= 450 \pm 37 \rm \mu Jy \ ,
\end{equation}
with $\chi^2/\rm dof=0.48$. The fit is presented in Figure \ref{Fig6}.

The errors quoted are statistical errors. The actual errors are
probably larger due to uncertainty in the radio flux due to
scintillation, and due to the fact that not all the observations are
exactly simultaneous (see Shepherd 1998).

Substituting these results in equations \ref{nu_a} and \ref{F_extr},
respectively, we obtain two constraints on the physical parameters of
the burst. The luminosity distance depends on the cosmological model.
For $\Omega=1$, $\Lambda=0$ (which implies:
$d_L=2(1+z-\sqrt{1+z})c/H_0$), $H_0=65 \rm Km s^{-1} Mpc^{-1}$ and
$z=0.835$ we find:
\begin{eqnarray}
\label{constraints2}
{(p-1)^{8/5}\over(p-2)}\left({p+2\over3p+2}\right)^{3/5}
\epsilon_e^{-1}\epsilon_B^{1/5}E_{52}^{1/5}n_1^{3/5}&=&5.43 \ ,\\
\label{constraints3}
{(p-1)^{6/5}\over(p-2)}\left({p+2\over3p+2}\right)^{1/5}
\epsilon_e^{-1}\epsilon_B^{2/5}E_{52}^{9/10}n_1^{7/10}&=&1.22 \ .
\end{eqnarray}

Substitution of the value for $\nu_a$ from the data (equation
\ref{constraints1}) into equation \ref{nu_a}, yields an equation
different by a factor of $\sim 2$ from equation $22$ of Wijers \&
Galama (1998). The difference is mainly due to a different theoretical
value they used for $\nu_a$. They also used a slightly different value
of $\nu_a$ as corresponding to the same observational data. This
factor of $\sim 2$ implies significant corrections to the values of
the physical parameters of the burst (see the second row of Table 1).
For example, $n_1$ (or $n$ in their notation) becomes a factor of
$\sim 20$ larger: $n_1=0.7$ instead of $n_1=0.035$. 

\newcommand{\rb}[1]{\raisebox{1.5ex}[0pt]{#1}}
\begin{table}
\begin{center}{\bf TABLE {\large 1}\\ \vspace{0.5cm}
\bf  Estimates of  the Physical Parameters
 of GRB970508.\\}\vspace{0.5cm}
\begin{tabular}{|c||c|c|c|}\hline 
& broken & modified & modified $\nu_a$,  \\
\rb{model}  & power law & {$\nu_a$} & $\nu_m$,
$F_{\nu_m}$, $\nu_c$ \\ \hline \hline
$E_{52}$ & 3.7 & 2 & 0.53 \\ \hline 
$\epsilon_e$ & 0.13 & 0.24 & 0.57 \\ \hline
$\epsilon_B$ & 0.068 & 0.011 & 0.0082 \\ \hline
$n_1$ & 0.035 & 0.70 & 5.3 \\ \hline
\end{tabular}\\
\end{center}
\caption{The first row depicts the values of the physical parameters of
  GRG970508 as calculated by Wijers \& Galama (1998). The other rows
  show how these values change when some of the equations they used in 
  the calculation are corrected. The first line lists the measurable 
  quantities, whose equations were altered.}
\end{table}

We modify the estimates further using more detailed calculations of
the spectrum near the peak flux (GPS), and use a different estimate
for the cooling frequency $\nu_c$ (Sari, Piran \& Narayan 1998). We
discover that the values obtained for the physical parameters of a
burst are very sensitive to the theoretical model of the spectrum. In
order to illustrate this we show in Table 1 the physical parameters of
GRB970508 for different estimates of the spectrum. The first column
uses a broken power law spectrum (Wijers \& Galama 1998); the second
column uses a corrected theoretical value for $\nu_a$ from equation
\ref{nu_a} and a corrected observational value for $\nu_a$ from
equation \ref{constraints1}; the third column adds a modified
theoretical values to $\nu_m$ and $F_{\nu_m}$ (which are taken from
GPS - denoted there as $\nu_{peak}$ and $F_{\nu,max}$) and a different
estimate for the cooling frequency $\nu_c$ (from Sari, Piran \&
Narayan 1998), keeping the observational values from Wijers \& Galama
(1998).

Our best values (with all modifications added) are $E_{52}= 0.53$,
$\epsilon_e = 0.57$, $\epsilon_B = 0.0082$ and $n_1 = 5.3$. These
values differ by one to two orders of magnitude than the values
obtained by Wijers \& Galama (1998) using a broken power law spectra.
One must keep in mind that it is difficult to obtain an accurate
estimate of the various observables from the observed data. It is
especially difficult to determine $\nu_m$, $F_{\nu_m}$ and $\nu_c$.
Therefore, more than obtaining a better estimate of the physical
parameters of the burst, these calculations show the sensitivity of
this method and the need for more accurate data.

Sufficient observational data has been gathered on the radio afterglow
of GRB980329, to enable a fit similar to the one we made for
GRB970508. Such a fit was carried out by Taylor et al (1998).  The
theoretical formula for the flux density that was used for the fit is
identical to equation \ref{disk}. Since this is a good approximation
for the shape of the spectra near $\nu_a$, we can use the values
extracted from the data to obtain constraints on the physical
parameters of GRB980329.
  
The values used by Taylor et al for the flux density are mean values
over the first month. Since $F_{\nu_a} \propto T^{1/2}$, this
corresponds to the flux density at $\sim 15 \ \rm days$. The values
extracted from the data are:
\begin{equation}
\label{constraints4}
\nu_a\cong1.3 \times 10^{10} \rm Hz \quad,\quad 
F_{\nu_a,ext}\cong 600 \rm \mu Jy \ .
\end{equation}
Substituting these results in equations \ref{nu_a} and \ref{F_extr},
respectively, we obtain two constraints on the parameters of
GRB980329. For $\Omega=1$, $\Lambda=0$ and $H_0=65 \rm Km s^{-1}
Mpc^{-1}$ we find:
\begin{eqnarray}
\label{constraints5}
{2\over(1+z)}{(p-1)^{8/5}\over(p-2)}\left({p+2\over3p+2}\right)^{3/5}
\epsilon_e^{-1}\epsilon_B^{1/5}E_{52}^{1/5}n_1^{3/5}&\cong&26 \ ,\\
\label{constraints6}
\left({\sqrt2-1\over\sqrt{1+z}-1}\right)^2
{(p-1)^{6/5}\over(p-2)}\left({p+2\over3p+2}\right)^{1/5}
\epsilon_e^{-1}\epsilon_B^{2/5}E_{52}^{9/10}n_1^{7/10}&\cong&1.5 \ ,
\end{eqnarray}
where the red shift $z$ of this burst is not yet known.

\section{Discussion}

We have considered synchrotron emission from a system moving
relativistically, taking into account the effect of synchrotron self
absorption. We have assumed an adiabatic evolution and used the self
similar solution of Blandford \& McKee (1976) to describe the matter
behind the shock. This solution describes an extreme relativistic
spherical blast wave expanding into a cold uniform medium. Our
calculations accounted for the emission from the whole region behind
the shock front.

We have assumed a power law distribution of electrons with an
isotropic velocity distribution. Three alternative models have been
considered for the evolution of the magnetic field, including an
equipartition model. We have calculated the flux density at
frequencies near the self absorption frequency $\nu_a$, under the
assumption that $\nu_a \ll \nu_m$.

We have obtained an expression for the self absorption frequency
$\nu_a$, which we defined as the frequency at the point where the
extrapolations of the asymptotic power laws above and below the
$\nu_a$ meet. The value we obtained for $\nu_a$ is close to the value
obtained by Wijers \& Galama (1998), and a factor of $\sim 4$ for
$p=2.5$ (or a factor of $\sim 7$ for $p=2.2$) larger than the value
obtained by Waxman (1997b).

We have calculated the observed spectra near $\nu_a$ for three
different magnetic field models (Figure \ref{Fig4}). The spectra
differs more than a few percent from the asymptotic power laws only
for frequencies less than half an order of magnitude above or below
$\nu_a$. This applies to all the magnetic field models we have
considered.

Together with the afterglow images obtained in GPS we now have a
complete set of the observed image over a wide range of frequencies
(see Figures \ref{Fig4} and \ref{Fig5}). For $\nu \gg \nu_m$ there is
a thin bright ring on the outer edge of the image, and the surface
brightness at the center of the image is only a few percent of its
maximal value. For $\nu_a \ll \nu < \nu_m$ there is a wider ring and
the contrast between the center and the edge of the image is smaller,
with $58\%$ of the average surface brightness at the center. For $\nu
\ll \nu_a$ the image is even more homogeneous, with $77\%$ of the
average surface brightness at the center.

The observed image in the radio frequencies is of special importance,
since the best resolution is obtained in radio, with VLBI.  A
sufficient resolution could be reached with a nearby GRB ($z \sim
0.2$) to resolve the inner structure of the image in radio frequencies.

We have fitted the theoretical spectra to observational data of the
afterglow of GRB970508 (Shepherd 1998) and extracted the values of
$\nu_a$ and $F_{\nu_a,ext}$ from the data (equation
\ref{constraints1}). Substituting these values into the theoretical
expressions we obtained two equations for the physical parameters. The
behavior of the spectra near $\nu_m$ can supply two other equations,
namely the equations for the peak frequency and for the peak flux
(adding only one independent equation to the two equations for $\nu_a$
and $F_{\nu_a,ext}$). An additional independent equation is obtained
for $\nu_c$. The power law $p$ of the electron distribution can be
determined from the high energy slope ($\nu \gg \nu_m$). It is
therefore possible with sufficient observational data to determine all
the physical parameters of the afterglow: $p$, $\epsilon_e$,
$\epsilon_B$, $E_{52}$ and $n_1$. A similar calculation (using an
equation for the cooling frequency $\nu_c$ instead of $F_{\nu_a,ext}$)
was made by Wijers \& Galama (1998) for GRB970508.  In fact combining
this equation we could even have an over-constrained system and check
for consistency of the solution.  The equation we obtained from
$\nu_a$ differs by a factor of $\sim 2$ from the corresponding
equation given in Wijers \& Galama.  This has a significant effect (up
to a factor of $\sim 20$) on the values of the physical parameters of
the burst which they have given.  We have shown (Table 1) that if
other equations are corrected as well, the values of the physical
parameters vary even more (up to two orders of magnitude).

\vspace{1cm} We thank Ehud Cohen for useful discussions and Dale Frail
and Shri Kulkarni for observational information. This research was
supported by NASA Grant NAG5-3516, and a US-Israel Grant 95-328.
Re'em Sari thanks The Clore Foundation for support.

\end{document}